\documentclass{article-hermes}
\usepackage[dvips]{graphicx}
\usepackage{subfigure}

\begin{document}

\proceedings{BDA 2003}{-}

\title[Une plate-forme dynamique...]{Une plate-forme dynamique pour l'évaluation des performances des bases de données à objets}

\author{Zhen He\fup{*} \andauthor  Jérôme Darmont\fup{**}}

\address{
\fup{*}Department of Computer Science\\
University of Vermont\\
Burlington, VT 05405, USA\\[3pt]
zhenhe@emba.uvm.edu\\[6pt]
\fup{**}ERIC, Université Lumière Lyon 2\\
5 avenue Pierre Mendès-France\\
69676 Bron Cedex, France\\[3pt]
jerome.darmont@univ-lyon2.fr
}

\resume{Dans les bases de données orientées-objets ou relationnelles-objets telles que les bases de données
multimédia et la plupart des bases de données XML, les séquences d'accès aux objets ne sont pas statiques. En effet,
les applications n'accèdent pas systématiquement aux mêmes objets dans le même ordre. C'est néanmoins
en utilisant de telles séquences d'accès que les performances de ces systèmes de gestion de bases de données et de techniques d'optimisation
associées telles que le regroupement ont été évaluées jusqu'ici. Cet article propose 
une plate-forme dynamique pour l'évaluation des performances des bases de données à objets baptisée DOEF.
DOEF permet des évolutions de séquence d'accès par la définition de styles d'accès paramétrables. Ce
premier prototype a été conçu pour être ouvert et extensible. Bien qu'à l'origine élaboré pour le
modèle orienté-objet, il peut également être utilisé dans un contexte relationnel-objet moyennant 
quelques adaptations. De plus, de nouveaux modèles d'évolution de séquence d'accès peuvent également y être inclus
facilement.\\
Pour illustrer les possibilités offertes par DOEF, nous avons mené deux séries \linebreak d'expérimentations.
Dans la première, nous avons utilisé DOEF pour comparer les performances de quatre algorithmes de 
regroupement d'objets dynamiques. Les résultats ont montré que DOEF pouvait effectivement évaluer
l'adaptabilité de chaque technique à différentes évolutions de séquence d'accès. Ils nous ont également
permis de conclure que les algorithmes de regroupement dynamique actuels tolèrent des évolutions de 
séquence d'accès lentes, mais que leurs performances se dégradent drastiquement lorsque ces évolutions
sont rapides. Dans notre seconde série de tests, nous avons utilisé DOEF pour comparer les performances
de deux gestionnaires d'objets persistants : Platypus et SHORE. Nos résultats ont permis de mettre en
évidence une défaillance de Platypus au niveau de la pagination disque.}

\abstract{In object-oriented or object-relational databases such as multimedia databases or most XML databases, access patterns are not 
static, i.e., applications do not always access the same objects in the same order repeatedly.  However, this has been the way these 
databases and associated optimisation techniques such as clustering have been evaluated up to now.  
This paper opens up research regarding this issue by proposing a dynamic object evaluation framework (DOEF).  DOEF accomplishes access pattern 
change by defining configurable styles of change. It is a preliminary prototype that has been designed to be open and fully extensible.  
Though originally designed for the object-oriented model, it can also be used within the object-relational model with few adaptations. 
Furthermore, new access pattern change models can be added too.\\
To illustrate the capabilities of DOEF, we conducted two different sets of experiments.  In the first set of experiments, we used DOEF to compare the performances of four 
state of the art dynamic clustering algorithms.  The results show that DOEF is effective at determining the 
adaptability of each dynamic clustering algorithm to changes in access pattern.  They also  
led us to conclude that dynamic clustering algorithms can cope with moderate levels of access pattern change, but that performance rapidly 
degrades to be worse than no clustering when vigorous styles of access pattern change are applied.  In the second set of experiments, 
we used DOEF to compare the performance of two different object stores: Platypus and SHORE.  
The use of DOEF exposed the poor swapping performance of Platypus.  
}

\motscles{Bases de données orientées-objets et relationnelles-objets,
Bancs d'essais, 
Evaluation de performance,
Regroupement.}


\maketitlepage

\section{Introduction}

L'évaluation de performance est une tâche critique à la fois pour les concepteurs de Systèmes de 
Gestion de Bases de Données à Objets\footnote{Dans la suite de cet article, nous utilisons les termes
Systèmes de Gestion de Bases de Données à Objets (SGBDO) pour désigner indifféremment les systèmes 
orientés-objets et relationnels-objets. La plupart des SGBD multimédia et XML sont des SGBDO.}
(architecture ou optimisation) et leurs utilisateurs (comparaison de performance, optimisation).
Traditionnellement, ces évaluations s'effectuent à l'aide de bancs d'essais constitués de modèles
de charge synthétiques (bases de données et opérations) et de mesures de performance. Aucun des
bancs d'essais conçus pour les SGBDO ne permet de modifier en cours de test les séquences d'accès
aux objets. Cependant, dans la réalité, la plupart des applications n'accèdent pas de manière 
répétitive au même ensemble d'objets, dans le même ordre. Par ailleurs, aucune des nombreuses études
concernant le regroupement (\emph{clustering}) dynamique d'objets ne contient d'indication sur la manière dont les
algorithmes proposés réagissent dans un tel environnement dynamique.

La capacité d'adaptation aux évolutions de séquence d'accès est pourtant critique pour obtenir
de bonnes performances. Optimiser une base de données pour bien répondre à une séquence d'accès
particulière peut d'ailleurs entraîner des dégradations de performance importantes lorsques d'autres
séquences d'accès sont employées. De plus, l'étude des performances d'un système sur une seule trace donnée
fournit peu d'indications aux concepteurs d'un système, qui ont besoin de cerner et d'optimiser le
comportement des composants de leur système dans différents cas d'utilisation. En opposition aux bancs d'essais du TPC~\cite{TPC},
qui proposent des outils d'évaluation {\em standardisés} permettant aux vendeurs et aux clients de
comparer des systèmes, l'objectif de notre plate-forme DOEF ({\em Dynamic Object Evaluation Framework})
est de leur permettre d'explorer les performances d'un SGBDO pour {\em différents} modèles d'évolution de séquence d'accès aux données.

DOEF est une première tentative de modélisation du comportement dynamique d'une application. La plate-forme
décrit un ensemble de protocoles qui définissent à leur tour un ensemble de modèles d'évolution de séquence d'accès.
DOEF a été conçu comme une surcouche logicielle d'OCB~\cite{OCBJournal},
qui est un banc d'essais générique capable de simuler le comportement des principaux bancs d'essais orientés-objets.
DOEF réutilise la base de données très riche et les opérations d'OCB. En tant
que surcouche, DOEF est facile à ajouter sur toute implémentation existante d'OCB.

DOEF ne comprend certainement pas tous les styles d'accès possibles. Cependant, nous avons conçu notre plate-forme pour 
être totalement extensible. D'une part, il est facile d'y inclure de nouveau modèles d'évolution de séquence 
d'accès. D'autre part, le modèle générique d'OCB peut être implémenté dans un système relationnel-objet
et la plupart de ses opérations demeurent pertinentes. DOEF peut donc
également être utilisé dans un contexte relationnel-objet.

Pour illustrer les possibilités offertes par DOEF, nous avons dans un premier temps évalué les performances 
de quatre algorithmes de regroupement dynamique d'objets, et ce pour trois raisons. Premièrement, le
regroupement d'objet est une technique d'optimisation des performances efficace~\cite{Gerlhof1996}.
Deuxièmement, les performances des algorithmes de regroupement dynamique d'objets sont très sensibles
aux évolutions de séquence d'accès. Troisièmement, malgré cela, les performances de ces techniques
n'ont jamais été évaluées dans cette optique. Finalement, afin de tester l'efficacité de DOEF pour
évaluer les performances de systèmes réels, nous avons également comparé 
les performances de deux gestionnaires d'objets persistants existants.

Cet article est organisé comme suit. La Section~\ref{art} décrit brièvement les bancs d'essais pour 
SGBDO. La Section~\ref{ocb} présente le banc d'essais OCB. La Section~\ref{doef} détaille les
spécifications de la plate-forme DOEF. Nous présentons les résultats expérimentaux que nous avons
obtenus dans la Section~\ref{exp}. Finalement, nous concluons cet article et présentons quelques
perspectives de recherche dans la Section~\ref{conclu}.

\section{Bancs d'essais existants}
\label{art}

Nous décrivons ici les principaux bancs d'essais pour SGBDO, hormis OCB, qui ont été proposés dans la littérature.
Il est important de noter qu'aucun d'eux ne présente de comportement dynamique.

Les bancs d'essais orientés-objets, dont les principaux sont OO1~\cite{OO1}, HyperModel~\cite{hypermodel},
OO7~\cite{OO7Benchmark} et Justitia~\cite{Justitia}, ont tous été conçus pour modéliser des applications
d'ingénierie (CAO, AGL, etc.). Leur spectre s'étend d'OO1, qui présente un schéma très simple (deux classes) et 
seulement trois opérations, à OO7, qui est plus générique et fournit un schéma complexe et 
paramétrable (dix classes), ainsi qu'une gamme d'opérations étendue (quinze opérations complexes). 
Justitia présente la particularité de prendre en compte des utilisateurs multiples, ce qui lui permet
de mieux modéliser des applications de type client-serveur. Néanmoins, même le schéma d'OO7 demeure 
statique et n'est pas suffisamment générique pour modéliser d'autres types d'applications que des
applications d'ingénierie (comme des applications financières,
multimédia, ou de télécommunication, par exemple~\cite{OO7Eval}). De plus, chaque pas vers une plus
grande complexité a rendu ces bancs d'essais de plus en plus difficiles à implémenter.

Les bancs d'essais relationnels-objets, tels que BUCKY~\cite{bucky} et BORD \linebreak
\cite{bord}, 
sont orientés-requêtes et uniquement dédiés aux
systèmes relationnels-\linebreak objets. Par exemple, BUCKY ne propose que des opérations spécifiques de ces
systèmes, considérant que la navigation typique parmi des objets est déjà traitée par d'autres bancs d'essais (voir ci-dessus).
Ces bancs d'essais se concentrent donc sur des requêtes impliquant des identifiants d'objets, l'héritage,
les jointures, les références de classes et d'objets, les attributs multivalués, l'"applatissement" des
requêtes, les méthodes d'objets et les types de données abstraits. Les schémas des bases de données de ces
bancs d'essais sont également statiques.

Finalement, des ensembles de charges ont été proposés pour mesurer les performances de SGBDO clients-serveurs~\cite{carey91,franklin93}.
Ces charges opèrent sur des pages plutôt que sur des objets. La notion de région chaude ou froide
(certaines "zones" de la base de données sont plus fréquemment accédées que d'autres) est également
avancée afin de modéliser le comportement d'applications réelles. Cependant, la région chaude ne
varie jamais. Le modèle de comportement n'est donc pas dynamique.

\section{Banc d'essais OCB}
\label{ocb}

OCB ({\em Object Clustering Benchmark}) est un banc d'essais générique et paramétrable destiné à évaluer les performances des SGBD
orientés-objets~\cite{OCBJournal}. Dans un premier temps spécialisé dans l'évaluation des stratégies
de regroupement, il a été étendu par la suite pour devenir totalement générique. Sa flexibilité et
son adaptabilité sont obtenues à l'aide d'un ensemble complet de paramètres. OCB est capable de 
simuler le fonctionnement des bancs d'essais orientés-objets qui constituent des standards de fait
(OO1, HyperModel et OO7). De plus, le modèle générique d'OCB peut être implémenté au sein d'un SGBD
relationnel-objet et la plupart de ses opérations y demeurent pertinentes. Nous ne proposons ici qu'un survol 
d'OCB. Ses spécifications complètes sont disponibles dans~\cite{OCBJournal}. Les deux composants 
principaux d'OCB sont sa base données et ses opérations.

\subsection{Base de données}

\begin{figure*}[hbt]
  \begin{center}
  \includegraphics[width = 10 cm]{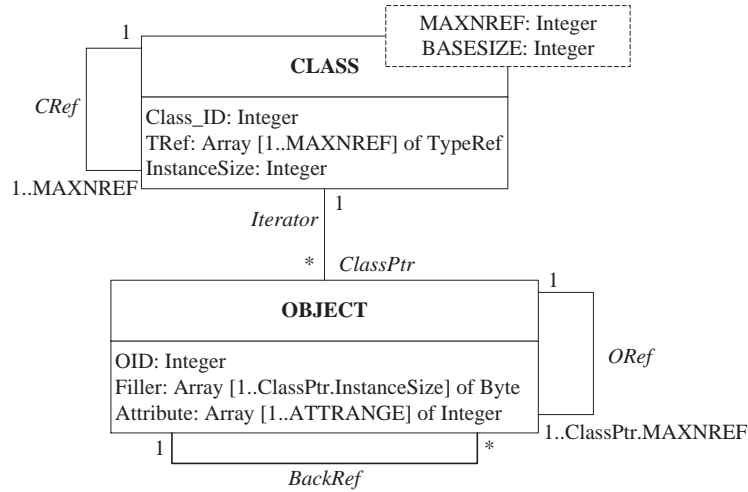}
  \caption{Schéma de la base de données d'OCB}
  \label{schema_ocb}
  \end{center}
\end{figure*}

Le schéma de la base de données d'OCB est constitué de {\em NC} classes
dérivées d'une même métaclasse (Figure~\ref{schema_ocb}). Les classes sont définies par deux paramètres :
{\em MAXNREF}, le nombre maximum de références à d'autres classes et {\em BASESIZE}, un incrément
utilisé pour calculer la taille des instances ({\em InstanceSize}).
Chaque référence de classe ({\em CRef}) possède un type {\em TRef}. Il existe {\em NTREF} différents
types de références (héritage, aggrégation...). Finalement, un itérateur ({\em Iterator}) est
maintenu au sein de chaque classe pour sauvegarder les références à toutes ses instances.
Chaque objet possède {\em ATTRANGE} attributs entiers qui peuvent être lus et mis à jour par les
opérations. Une chaîne de caractères ({\em Filler}) de taille {\em InstanceSize} permet de simuler
la taille réelle de l'objet.
Après instantiation du schéma, un objet  {\em O} de classe {\em C} pointe à travers les références
{\em ORef} vers au plus {\em C.MAXNREF} objets. Il existe également une référence inverse
({\em BackRef}) qui lie chaque objet référencé à celui qui le référence.
Les principaux paramètres qui définissent la base de données sont résumés dans le
Tableau~\ref{param_bd}.

\begin{table*}[hbt]
 \begin{center}
   \begin{tabular}{|l|l|c|} \hline
   {\bf Nom} &	{\bf Paramètre} &	{\bf Valeur par défaut}\\\hline\hline
NC &	Nombre de classes dans la BD &	50\\\hline
MAXNREF({\em i}) &	Nombre maxi de références par classe &	10\\\hline
BASESIZE({\em i}) &	Taille de base des instances, par classe &	50 octets\\\hline
NO &	Nombre total d'objets &	20000\\\hline
NREFT &	Nombre de types de références &	4\\\hline
ATTRANGE &	Nombre d'attributs d'un objet &	1\\\hline
CLOCREF &	Localité de réference de classe &	NC\\\hline
OLOCREF &	Localité de réference d'objet &	NO\\\hline
   \end{tabular}
   \caption{Principaux paramètres de la base de données d'OCB}
   \label{param_bd}
 \end{center}
\end{table*} 

\subsection{Opérations}

Les opérations d'OCB sont subdivisées en quatre catégories.

\begin{itemize}

\item \emph{Accès aléatoire :}  Accès à \emph{NRND} objets sélectionnés aléatoirement.

\item \emph{Balayage séquentiel :} Sélection aléatoire d'une classe et accès à toutes ses
instances. Une \emph{recherche par intervalle} effectue en plus un test sur la valeur de \emph{NTEST} 
attributs, pour chaque instance lue.

\item \emph{Parcours :} Il existe deux types de parcours.
Les \emph{accès ensemblistes} (ou associatifs) effectuent un parcours en largeur d'abord.
Les \emph{accès navigationnels} sont subdivisés en \emph{parcours simples} (en profondeur
d'abord), en \emph{parcours hiérarchiques} qui suivent toujours le même type de référence, et en 
\emph{parcours stochastiques} qui sélectionnent le lien à déréférencer de manière aléatoire.  
Chaque parcours démarre d'un objet racine sélectionné aléatoirement et procède jusqu'à une
profondeur prédéterminée. Tous les parcours peuvent être inversés pour suivre les liens {\em BackRef}.

\item \emph{Mise à jour:} Il existe trois types de mises à jour.
Les \emph{évolutions de schéma} et les \emph{évolutions de base de données} sont des insertions et des suppressions aléatoires de classes
et d'objets, respectivement.
Les \emph{mises à jour d'attributs} sélectionnent aléatoirement \emph{NUPDT} objets 
à mettre à jour ou bien sélectionnent aléatoirement une classe et mettent à jour toutes
ses instances (\emph{mise à jour séquentielle}).  

\end{itemize}

\section{Spécification de DOEF}
\label{doef}

\subsection{Contexte dynamique}

Afin d'illustrer notre propos, nous commençons par donner un exemple de scénario que notre plate-forme
peut simuler. Supposons que nous modélisons une librairie en ligne dans laquelle certains types de 
livres se vendent bien à des périodes données. Par exemple, les guides touristiques sur l'Australie ont
pu être populaires lors des Jeux Olympiques 2000. Mais une fois l'événement terminé, ces livres se sont
soudainement ou graduellement moins bien vendus. 

Une fois un livre sélectionné, des informations le concernant peuvent être demandées, comme par exemple
un résumé, une photo de la couverture, des extraits, des critiques,  etc. Dans un SGBDO, cette information est stockée
sous la forme d'objets référencés par l'objet (le livre) sélectionné. Accéder à ces informations revient
donc à naviguer dans un graphe d'objets dont la racine est l'objet initialement sélectionné (ici, un livre).
Après avoir consulté les informations concernant un livre, un utilisateur peut ensuite choisir un autre
ouvrage du même auteur, qui devient alors la racine d'un nouveau graphe de navigation.

Nous décrivons ici les cinq principales étapes de notre démarche et les illustrons à l'aide de l'exemple
que nous venons de présenter.

\begin{enumerate}

\item \textbf{Paramétrage des H-régions :} Nous divisons tout d'abord la base de données en régions à
probabilité d'accès homogène (appelées H-régions). Dans notre exemple, chaque H-région représente un
ensemble de livres différent, chacun de ces ensembles possédant une probabilité d'accès propre.

\item \textbf{Spécification de la charge :} Les H-régions permettent d'assigner des probabilités d'accès
aux objets, mais pas de déterminer les actions à effectuer une fois un objet sélectionné. Nous appelons
ces objets sélectionnés \emph{racines de la charge} ou simplement racines. Dans cette étape, nous sélectionnons
le type d'opération à effectuer sur une racine parmi ceux proposés dans OCB. Dans notre exemple, la charge est un
parcours de graphe d'objets qui va de l'ouvrage sélectionné à l'information requise (par exemple, un
extrait du livre).

\item \textbf{Spécification du protocole régional :} Les protocoles régionaux exploitent les H-régions 
pour accomplir l'évolution de séquence d'accès. Différents modèles d'évolution de séquence d'accès
peuvent être obtenus en faisant varier le paramétrage des H-régions au cours du temps. Par exemple, un
protocole régional peut tout d'abord affecter une grande probabilité d'accès à une H-région donnée, tandis
que les autres H-régions ont une faible probabilité d'accès. Après un certain temps, une H-région
différente hérite de la grande probabilité d'accès. Dans notre exemple de librairie en ligne, cela modélise
la baisse d'intérêt pour les guides touristiques sur l'Australie après la fin des Jeux Olympiques.

\item \textbf{Spécification du protocole de dépendance :} Les protocoles de dépendance nous permettent
de spécifier une relation entre la racine courante et la racine suivante. Dans notre exemple, cela peut modéliser
un client qui choisit de consulter un ouvrage écrit par le même auteur que l'ouvrage qu'il a sélectionné
précédemment.

\item \textbf{Intégration du protocole régional et du protocole de dépendance :} 
Dans cette étape, nous intégrons les protocoles régionaux et de dépendance afin de modéliser des évolutions
de dépendance entre des racines successives. Par exemple, un client de notre librairie en ligne peut
sélectionner un livre qui l'intéresse, puis découvrir une liste d'ouvrages du même auteur qui font partie
des meilleures ventes actuelles. Le client sélectionne alors l'un des livres (protocole de 
dépendance). L'ensemble des livres du même auteur, qui font partie
des meilleures ventes \emph{actuelles}, est lui susceptible de changer avec le temps (protocole régional).

\end{enumerate}

\subsection{H-régions}

Les H-régions sont des sous-ensembles de la base de données de probabilité d'accès homogène. Les paramètres
qui les définissent sont détaillés ci-dessous.

\begin{itemize}

\item \emph{HR\_SIZE} : Taille de la H-région (fraction de la taille de la base).

\item \emph{INIT\_PROB\_W} : Poids initial de la H-région. La probabilité d'accès est égale
à ce poids divisé par la somme des poids de toutes les H-régions.

\item \emph{LOWEST\_PROB\_W} : Poids minimum pour la H-région.

\item \emph{HIGHEST\_PROB\_W} : Poids maximum pour la H-région.

\item \emph{PROB\_W\_INCR\_SIZE} : Incrément par lequel le poids est augmenté ou diminué lorsqu'une
évolution survient.

\item \emph{OBJECT\_ASSIGN\_METHOD} : Détermine la manière dont les objets sont assignés à une H-région.
La sélection \emph{aléatoire} permet de les choisir au hasard dans la base de données. La sélection
\emph{par classe} trie tout d'abord les objets selon leur identifiant de classe, avant de sélectionner les
$N$ premiers, $N$ étant le nombre d'objets alloués à la H-région.

\item \emph{INIT\_DIR} : Direction initiale dans laquelle évolue l'incrément de poids (haut ou bas).
      
\end{itemize}

\subsection{Protocoles régionaux}

Les protocoles régionaux simulent les évolutions de séquence d'accès en initialisant tout d'abord les paramètres
de toutes les H-régions. Ces paramètres sont ensuite modifiés périodiquement de manière prédéterminée.
Cet article présente trois modèles d'évolution régionale.

\subsubsection{Fenêtre mobile}

Ce protocole régional simule des changements brusques dans la séquence d'accès. Dans notre exemple, cela
correspond à un livre qui devient populaire tout d'un coup (suite à une promotion TV, par exemple).
Une fois l'événement passé, l'ouvrage retrouve sa cote de popularité initiale très rapidement. Ce modèle
d'évolution est obtenu en déplaçant une fenêtre dans la base de données. Les objets de la fenêtre 
présentent une probabilité d'être sélectionnés comme racine très supérieure à celle des autres objets de la base.
Nous atteignons cet objectif en divisant la base de données en $N$ H-régions de tailles égales. Une de
ces H-régions est choisie pour être la première région "chaude" (celle qui a la plus haute probabilité
d'accès). Après un certain nombre de sélections de racine successives, une nouvelle H-région devient la
région chaude.

\begin{itemize}

\item La base de données est divisée en \emph{N} H-régions de tailles égales.

\item Le paramètre \emph{INIT\_PROB\_W} d'une H-région est fixé à
\emph{HIGHEST\_PROB\_W} (région chaude). La valeur de
\emph{INIT\_PROB\_W} pour les autres H-régions est fixée à \emph{LOWEST\_PROB\_W}.  

\item Pour chaque H-région, \emph{PROB\_INCR\_SIZE} est égal à \emph{HIGHEST\_PROB\_W} - \emph{LOWEST\_PROB\_W}. 
Toutes les H-régions doivent avoir les mêmes \emph{LOWEST\_PROB\_W} et \emph{PROB\_W\_INCR\_SIZE}.

\item Soit un paramètre $H$ défini par l'utilisateur, qui reflète la vitesse d'évolution de séquence d'accès.

\item Le paramètre \emph{INIT\_DIR} de toutes les H-régions est fixé vers le bas. La fenêtre est au départ placée dans la région chaude.
Après 1 / \emph{H} sélections de racines, la fenêtre se déplace d'une H-région à une autre.  La région
qu'elle quitte a son paramètre \emph{INIT\_DIR} réinitialisé vers le bas, tandis que celle sur laquelle
elle arrive a son paramètre \emph{INIT\_DIR} réinitialisé vers le haut.
Les poids des H-régions sont ensuite incrémentés ou décrémentés, selon la valeur de \emph{INIT\_DIR}.

\end{itemize}

\subsubsection{Fenêtre mobile graduelle}

Ce protocole est similaire au précédent, mais la région chaude "se refroidit" graduellement et non brutalement.
Les régions froides "se réchauffent" également graduellement au fur et à mesure que la fenêtre passe au-dessus.
Cela permet de tester la faculté d'un système ou d'un algorithme à s'adapter à des types d'évolution plus doux. Dans notre exemple, ce modèle d'évolution peut décrire la baisse d'intérêt 
graduel pour les guides touristiques sur l'Australie après les Jeux Olympiques. Simultanément, les
guides touristiques pour d'autres destinations peuvent rencontrer plus de succès.

Ce protocole est défini de la même manière que le précédent, à deux exceptions près. Premièrement,
le paramètre \emph{PROB\_INCR\_SIZE} est fixé par l'utilisateur et non plus calculé. Sa valeur 
détermine l'intensité avec laquelle la séquence d'accès change à chaque itération. Deuxièmement,
l'évolution des probabilités d'accès à une H-région changent. Lorsque la fenêtre arrive sur une
H-région, la valeur de son paramètre \emph{INIT\_DIR} est inversée. Lorsque la fenêtre quitte une
H-région, elle reste en revanche inchangée. Cela permet à la H-région de continuer à se "réchauffer"
ou à se "refroidir" graduellement.

\subsubsection{Cycles d'évolution}

Ce modèle d'évolution décrit, par exemple, le comportement des clients d'une banque, qui tendraient à
appartenir à une catégorie (comportementale, socio-\linebreak professionnelle...) le matin et à une autre catégorie l'après-midi. Réitéré, ce processus 
crée un cycle d'évolution.

\begin{itemize}

\item La base de données est divisée en trois H-régions. Les deux premières représentent l'ensemble
d'objets qui subit le cycle, la dernière la partie de la base qui demeure inchangée. Les valeurs du paramètre
\emph{HR\_SIZE} des deux premières H-régions doivent être égales et sont spécifiées par l'utilisateur.
Celle de la troisième H-région correspond à la taille du reste de la base.

\item Les valeurs de \emph{LOWEST\_PROB\_W} et \emph{HIGHEST\_PROB\_W} pour les deux premières H-régions 
sont fixées de manière à refléter les valeurs extrêmes du cycle. 

\item La valeur de \emph{PROB\_INCR\_SIZE} est égale à \emph{HIGHEST\_PROB\_W} - \emph{LOWEST\_PROB\_W}
pour les deux premières H-régions, à zéro pour la dernière.

\item La valeur de \emph{INIT\_PROB\_W} est fixée à \emph{HIGHEST\_PROB\_W} pour la première H-région
et à  \emph{LOWEST\_PROB\_W} pour la deuxième.

\item La valeur de \emph{INIT\_DIR} est dirigée vers le bas pour la région "chaude" et vers le haut pour la région "froide". 

\item Un paramètre \emph{H} est de nouveau employé pour faire varier la vitesse d'évolution des séquences d'accès.

\end{itemize}

\subsection{Protocoles de dépendance}

Il existe de nombreux scénarios au cours desquels une personne exécute une requête, puis décide d'en 
exécuter une autre en se basant sur les résultats de la première, établissant ainsi une dépendance
entre les deux requêtes. Nous avons spécifié dans cet article quatre protocoles de dépendance.

\subsubsection{Sélection aléatoire}

Cette méthode utilise simplement une fonction aléatoire pour sélectionner la racine courante. Ce protocole
modélise une personne qui lance une toute nouvelle requête après avoir exécuté la précédente.

$r_i$ = $RAND\emph{1}()$, où $r_i$ est l'identifiant du $i^{eme}$ objet racine.  
La fonction $RAND\emph{1}()$ ne suit pas nécessairement une loi uniforme.

\subsubsection{Sélection par référence}

La racine courante est choisie parmi les objets référencés par la racine précédente. Dans notre exemple, 
ce scénario correspond à une personne qui termine la consultation d'un livre, puis recherche l'ouvrage
suivant dans la même série ou collection.

$r_{i+1}$ = $RAND\emph{2}(RefSet(r_{i},D))$, où $RefSet(r_i,D)$ est une fonction qui retourne
l'ensemble des objets référencés par 
la $i^{eme}$ racine.  
Nous avons utilisé deux types de références. Les références structurelles (S-références) sont simplement
celles qui sont issues du graphe d'objets. Nous avons de plus introduit les D-références dans le seul but 
d'établir des dépendances entre les racines de parcours. Le paramètre $D$ est utilisé pour spécifier
le nombre de D-références par objet.

\subsubsection{Sélection par objets parcourus}

La racine courante est sélectionnée dans l'ensemble d'objets retournés par la requête précédente. Par
exemple, un client demande une liste de livres avec leurs auteurs et leurs éditeurs, puis décide ensuite
de lire un extrait de l'un des ouvrages.

$r_{i+1}$ = $RAND\emph{3}(TraversedSet(r_i,C))$,
où $TraversedSet(r_i,C)$ retourne l'ensemble des objets référencés pendant le parcours effectué à
partir de la $i^{eme}$ racine.  Le paramètre $C$ sert à limiter le nombre d'objets retournés par
$TraversedSet(r_i,C)$.  C'est une fraction de la cardinalité de l'ensemble des objets retournés.  
Ainsi, le degré de localité des objets renvoyés par $TraversedSet(r_i,C)$ 
peut être contrôlé (plus $C$ est petit, plus le degré de localité est grand).

\subsubsection{Sélection par classe}

La racine courante doit appartenir à la même classe que la racine précédente. De plus, la sélection
de la racine est réduite à un sous-ensemble des objets de cette classe qui dépend de la racine
précédente. Par exemple, un client de notre librairie en ligne choisit un livre du même auteur
que l'ouvrage qu'il vient de consulter. Dans ce cas, la fonction de sélection par classe retourne
les livres écrits par le même auteur.

$r_{i+1}$ = $RAND\emph{4}(f(r_{i}, Class(r_{i}), U))$, où
$Class(r_i)$ retourne la classe de la $i^{eme}$ racine. Le paramètre $U$ est défini par l'utilisateur
et indique la cardinalité de l'ensemble d'objets retourné par la fonction $f()$. C'est en fait une fraction
de la cardinalité totale de la classe. Il peut être utilisé pour faire varier le degré de localité
entre les objets retournés par $f()$.  $f()$ doit être injective.

\subsubsection{Sélection hybride}
\label{selhyb}

Ce type de sélection permet d'utiliser une combinaison des protocles de dépendance décrits ci-dessus.
Cette possibilité est importante, car elle peut simuler par exemple un utilisateur qui initie une
requête totalement nouvelle après avoir suivi des dépendances. La sélection hybride est implémentée
en deux phases. La première \emph{phase aléatoire} utilise la sélection aléatoire pour sélectionner
une première racine. Dans la seconde \emph{phase de dépendance}, un des protocoles de dépendance
ci-dessus est appliqué pour sélectionner la racine suivante. La seconde phase est répétée $R$ fois
avant que la première ne survienne à nouveau. Les deux phases sont réitérées de manière continue.

La probabilité de sélectionner un protocole de dépendance donné dans la \emph{phase de dépendance}
est spécifiée à l'aide des paramètres suivants : \emph{RANDOM\_DEP\_PROB} (sélection aléatoire), 
\emph{SREF\_DEP\_PROB} (sélection par référence sur les S-références), \emph{DREF\_DEP\_PROB} 
(sélection par référence sur les D-références), \emph{TRAVERSED\_DEP\linebreak \_PROB} (sélection par objets parcourus)
et \emph{CLASS\_DEP\_PROB} (sélection par classe).

\subsection{Intégration des protocoles régionaux et de dépendance}

Les protocoles de dépendance modélisent le comportement des utilisateurs. \linebreak Comme ce dernier peut changer
au cours du temps, ces protocoles doivent également être capables d'évoluer. En intégrant les protocoles
régionaux et de dépendance, nous parvenons à simuler des évolutions de dépendance entre des sélections
successives de racines. Ceci est facilement accompli en exploitant la propriété des protocoles de dépendance 
de retourner un ensemble d'objets racines candidats en fonction d'une racine précédente donnée. Jusqu'à
présent, la racine courante était sélectionnée dans cet ensemble à l'aide d'une fonction aléatoire. Plutôt
que de procéder ainsi, nous avons partitionné l'ensemble de racines candidates en H-régions et appliqué
les protocoles régionaux à ces H-régions.
Lorsque le protocole de dépendance "sélection par objets parcourus" est utilisé, la propriété suivante
doit être vérifiée : pour un même objet racine, le même ensemble d'objets doit être parcouru. Ainsi, une
racine donnée donne toujours lieu au même parcours.

\section{Résultats expérimentaux}
\label{exp}

\subsection{Regroupement dynamique}

Dans cette série de tests, nous avons utilisé DOEF pour comparer les performances de quatre algorithmes dynamiques
de regroupement d'objets parmi les plus récents : DSTC~\cite{ecoop96_bullat}, DRO~\cite{DRO}, OPCF-PRP et OPCF-GP~\cite{OPCF}.
L'objectif du regroupement est de placer automatiquement les objets qui sont susceptibles d'être utilisés
au même moment dans une même page disque, afin de minimiser les entrées/sorties. 

DSTC, qui est basé sur la
collecte de statistiques d'utilisation au volume contrôlé, entraîne néanmoins une
surcharge de regroupement élevée. DRO se base en partie sur DSTC, mais utilise une plus petite quantité
de statistiques et génère une surcharge beaucoup plus faible. Finalement, OPCF est une plate-forme qui
permet de transformer des algorithmes de regroupement statiques en algorithmes dynamiques. Elle a été 
appliquée sur des stratégies de partitionnement de graphe (GP) et de hiérarchisation de probabilités (PRP).

Le paramétrage de ces techniques de regroupement est indiqué dans le Tableau~\ref{param_clust}. Il permet,
dans chaque cas, d'aboutir au meilleur résultat avec l'algorithme concerné. Par manque de place, ces
paramètres ne sont pas décrits ici, mais ils sont complètement documentés dans les articles qui présentent
les travaux correspondants.

\begin{table}[hbt]
\begin{center}
\subfigure[DSTC]{
\begin{tabular}{|l|l|}\hline
{\bf Paramètre} & {\bf Valeur}\\\hline\hline
$n$ & 200\\\hline
$n_p$ & 1\\\hline
$p$ & 1000\\\hline
$T_{fa}$ & 1.0\\\hline
$T_{fe}$ & 1.0\\\hline
$T_{fc}$ & 1.0\\\hline
$w$ & 0.3\\\hline
\end{tabular}}\quad
\subfigure[DRO] {
\begin{tabular}{|l|l|}\hline
{\bf Paramètre} & {\bf Valeur}\\\hline\hline
MinUR & 0.001 \\\hline
MinLT & 2 \\\hline
PCRate & 0.02 \\\hline
MaxD &  1\\\hline
MaxDR & 0.2 \\\hline
MaxRR & 0.95 \\\hline
SUInd & true \\\hline
\end{tabular}}\quad
\subfigure[OPCF] {
\begin{tabular}{|l|l|l|}\hline
{\bf Paramètre} & {\bf Valeur}\\\hline\hline
N & 200 \\\hline
CBT & 0.1 \\\hline
NPA  & 50   \\\hline
NRI & 25  \\\hline
\end{tabular}}\quad
\caption{Paramétrage des algorithmes de regroupement dynamiques}
\label{param_clust}
\end{center}    
\end{table}

Nous avons mené nos expérimentations à l'aide de la plate-forme de simulation à événements discrets
VOODB~\cite{voodb}, qui permet d'évaluer les performances des SGBDO en général et des méthodes d'optimisation
comme le regroupement en particulier. Nous avons fait ce choix de la simulation pour deux raisons. 
Premièrement, cela nous a permis de développer et de tester rapidement plusieurs algorithmes de
regroupement dynamiques, alors que les études menées jusqu'ici en comparaient deux au plus.
Deuxièmement, il est relativement aisé de simuler de manière précise des entrées/sorties, qui sont
la mesure d'efficacité  principale des algorithmes de regroupement. Le paramétrage de VOODB pour 
ces expérimentations est indiqué dans le Tableau~\ref{param_tests}~(a).

Puisque DOEF utilise la base d'objets d'OCB et ses opérations, il est également important d'indiquer
le paramétrage d'OCB pour ces tests (Tableau~\ref{param_tests}~(b)). La taille des objets varie de 50
à 1600 octets, pour une moyenne de 233 octets. Nous avons utilisé une base de 100000 objets, soit
une taille de 23,3~Mo. Bien que ce soit une petite base, nous avons également utilisé un petit cache mémoire
(4~Mo) afin que le rapport taille de la base sur taille du cache demeure important. Les performances
des algorithmes de regroupement sont en effet plus sensibles à ce rapport qu'à la taille de la base
seule. Nous avons par ailleurs utilisé une seule opération d'OCB : le parcours simple (de pronfondeur 2),
car c'est la seule qui permet de toujours accéder au même ensemble d'objets à partir d'une racine
donnée. Cela établit un lien direct entre des variations de sélection des racines et des évolutions
de séquence d'accès. Lors de chaque test, nous avons exécuté 10000 transactions.

Finalement, les principaux paramètres de DOEF sont présentés dans le \linebreak Tableau~\ref{param_tests}~(c). 
La valeur de \emph{HR\_SIZE} permet de créer une région chaude qui représente 3~\% de la base.
Les valeurs de \emph{HIGHEST\_PROB\_W} et de \emph{LOWEST\_PROB\_W} permettent d'affecter une probabilité
d'accès de 80~\% à la région chaude et de 20~\% aux régions froides, globalement. Nous avons paramétré
DOEF ainsi afin de refléter le comportement typique d'une application~\cite{gray87,carey91,franklin93}.

\begin{table}[hbt]
  \begin{center}   
  \subfigure[VOODB]{
    \begin{tabular}{|l|l|}\hline
      {\bf Paramètre} & {\bf Valeur}\\\hline\hline
      Système & Centralisé \\\hline
      Taille page & 4 Ko\\\hline
      Taille cache & 4 Mo  \\\hline
      Gestion cache & LRU\\\hline
      Préchargement & --- \\\hline
      Nb serveurs & 1\\\hline
      Nb utilisateurs & 1\\\hline
            Placement initial & Séquentiel\\\hline
    \end{tabular}}\quad
   \subfigure[OCB]{
      \begin{tabular}{|l|l|} \hline
        {\bf Paramètre} & {\bf Valeur} \\\hline\hline
        Nb classes& 50\\\hline
        Nb références& 10\\ \hline
        Taille objets & 50\\\hline
        Nb objets & 100000\\\hline
        Nb types références & 4 \\\hline
        Dist. types réf. & Uniforme\\\hline
        Dist. réf. classes & Uniforme\\\hline
        Dist. objets ds classes & Uniforme\\\hline
        Dist. réf. objets & Uniforme\\\hline
    \end{tabular}} \quad
   \subfigure[DOEF]{ 
	\begin{tabular}{|l|l|} \hline
        {\bf Paramètre} & {\bf Valeur} \\\hline\hline
	  \emph{HR\_SIZE} & 0.003\\\hline
	  \emph{HIGHEST\_PROB\_W} & 0.80\\\hline
	  \emph{LOWEST\_PROB\_W} & 0.0006\\\hline	
	  \emph{PROB\_W\_INCR\_SIZE} & 0.02\\\hline
	  \emph{OBJECT\_ASSIGN\_METHOD} & Aléatoire\\\hline
        \end{tabular}}\quad
  \caption{Paramétrage de l'environnement de test}
  \label{param_tests}
\end{center}    
\end{table}

Pour cette série de tests, nous nous concentrons sur la faculté des algorithmes de regroupement
à s'adapter à des évolutions de séquence d'accès, et non sur leur performance pure. Tous les résultats
présentés ici sont exprimés en termes de nombre d'entrées/sorties total, c'est-à-dire de somme des
entrées/sorties nécessaires à l'exécution des transactions et du regroupement (surcharge). Dans les figures 
qui suivent, l'étiquette {\bf NC} indique une expérience étalon sans regroupement d'objets. 

\subsubsection{Application de protocoles régionaux}

Dans cette série de tests, nous avons utilisé les protocoles régionaux \emph{fenêtre mobile} et \emph{fenêtre mobile graduelle}
pour évaluer la capacité des algorithmes de regroupement à s'adapter à des évolutions de séquence d'accès.
Pour cela, nous avons fait varier le paramètre $H$ (vitesse de changement des séquences d'accès).

\begin{figure*}[hbt]
  \begin{center}
  \subfigure[Fenêtre mobile]{\includegraphics[width = 5.8 cm]{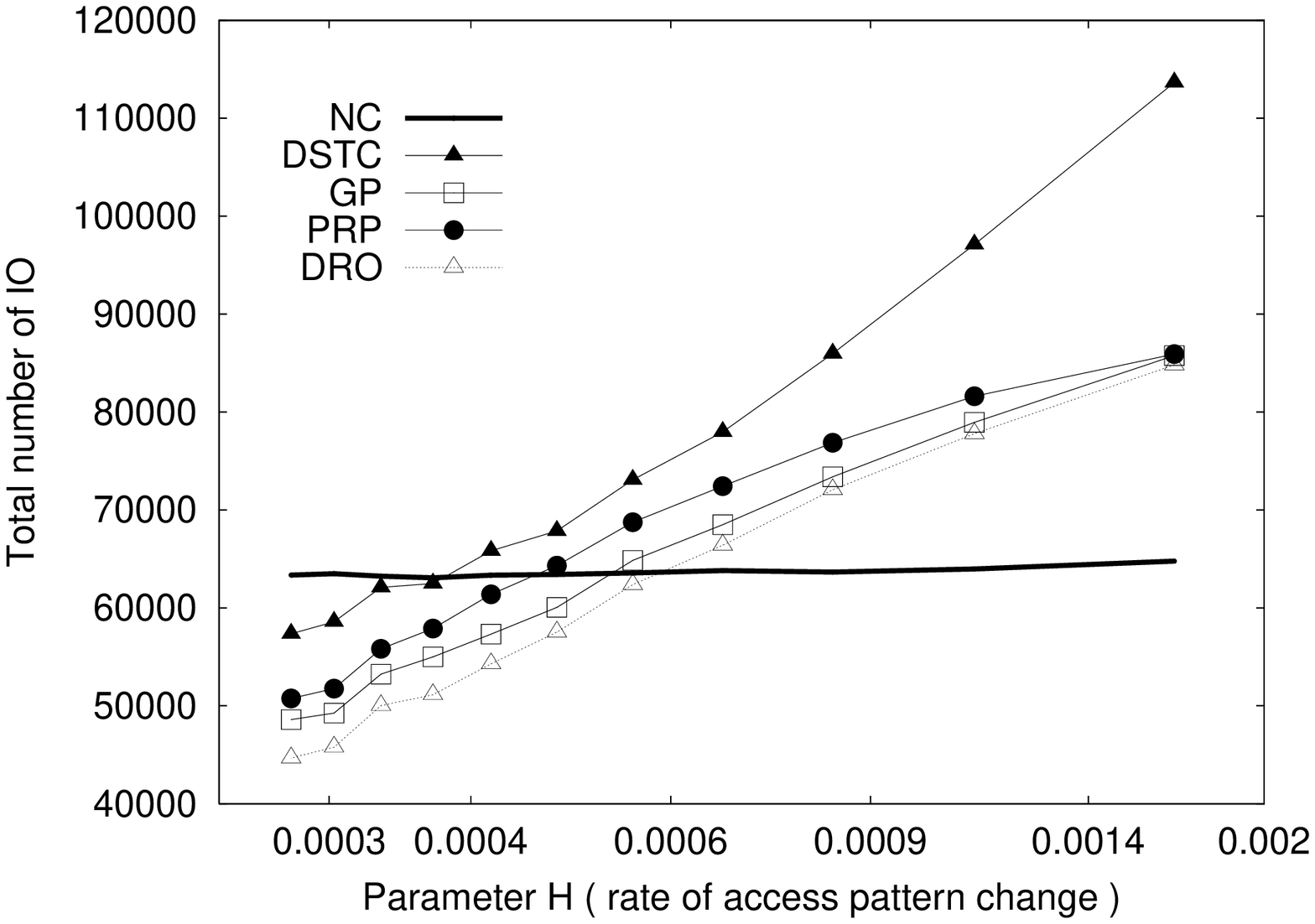}}\quad
  \subfigure[Fenêtre mobile graduelle]{\includegraphics[width = 5.8 cm]{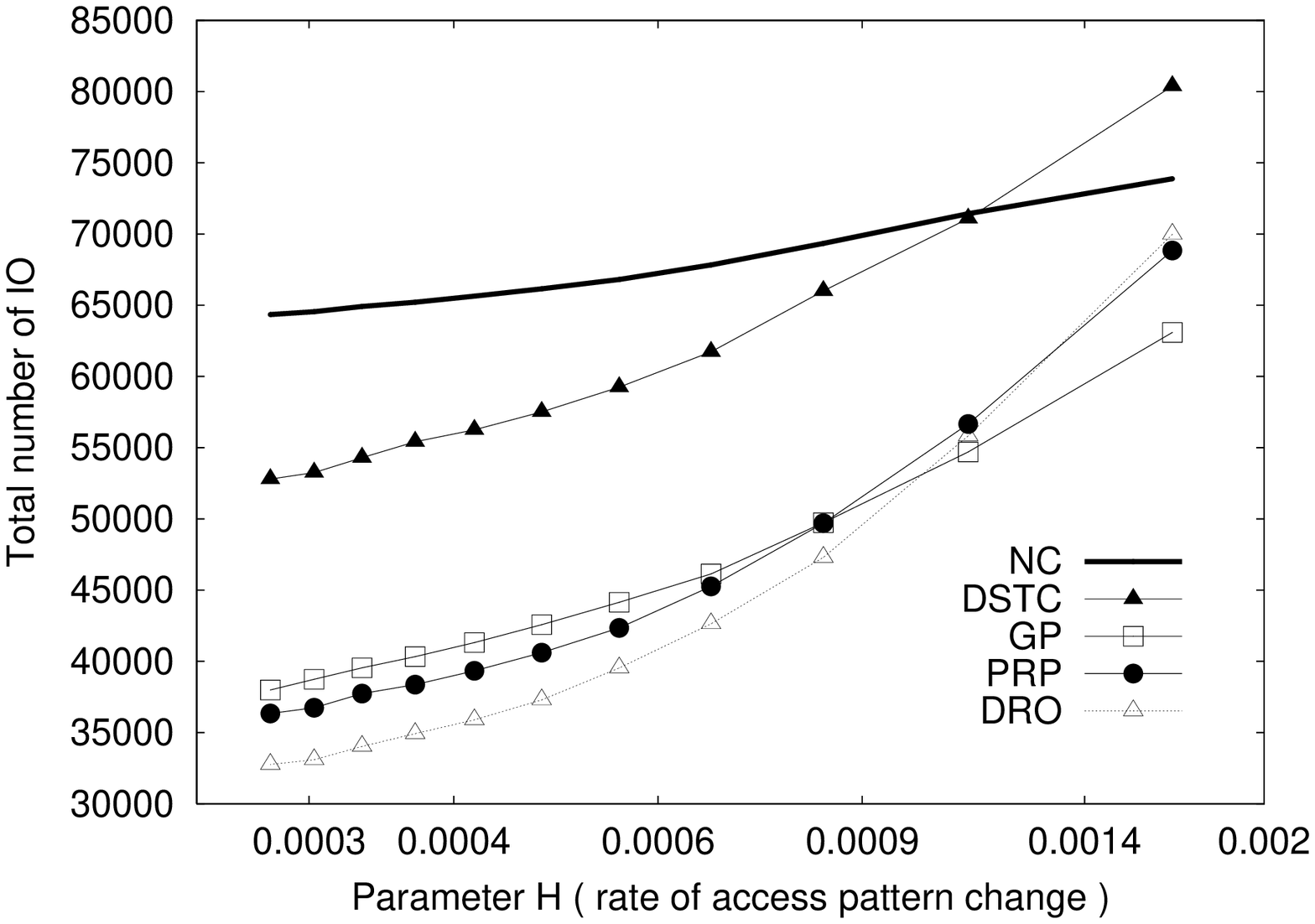}}\quad
  \caption{Résultat de l'application de protocoles régionaux}
  \label{res1}
  \end{center}  
\end{figure*}

Les résultats que nous avons obtenus sont représentés dans la Figure~\ref{res1}. Nous pouvons en tirer
trois conclusions. Premièrement, quand la vitesse de changement des séquences d'accès est faible ($H<0.0006$),
tous les algorithmes ont des performances similaires en termes de tendance. Cela signifie qu'ils réagissent
tous de manière analogue lorsque les séquences d'accès évoluent lentement. Deuxièmement, quand la vitesse
d'évolution est plus élevée (Figure~\ref{res1} (a)), leurs performances deviennent rapidement plus mauvaises que si aucun
regroupement n'était effectué (en raison de la surcharge qu'ils engendrent). Troisièmement, 
quand la vitesse de changement des séquences d'accès est élevée ($H>0.0006$), les 
algorithmes DRO, GP et PRP se montrent plus robustes que DSTC. Cela est dû au fait que ces
techniques ne réorganisent qu'un nombre limité de pages disque (celles qui sont mal regroupées).
Nous appelons cette propriété "regroupement prudent et flexible". Par contraste, DSTC peut déplacer
une page disque même quand le gain potentiel est faible. Ainsi, quand les séquences d'accès changent
rapidement, DSTC engendre une surcharge importante pour un bénéfice faible, ce qui explique ses moins
bonnes performances.

\subsubsection{Intégration de protocoles régionaux et de dépendance}

Dans cette expérience, nous avons exploré les effets d'une évolution de séquence d'accès sur le
protocole de dépendance \emph{sélection par S-référence} en l'intégrant avec les protocoles
régionaux \emph{fenêtre mobile} et \emph{fenêtre mobile graduelle}. La valeur du paramètre $R$ de cette
\emph{sélection hybride} a été fixée à 1.

\begin{figure*}[hbt]
  \begin{center}
  \subfigure[Fenêtre mobile]{\includegraphics[width = 5.8 cm]{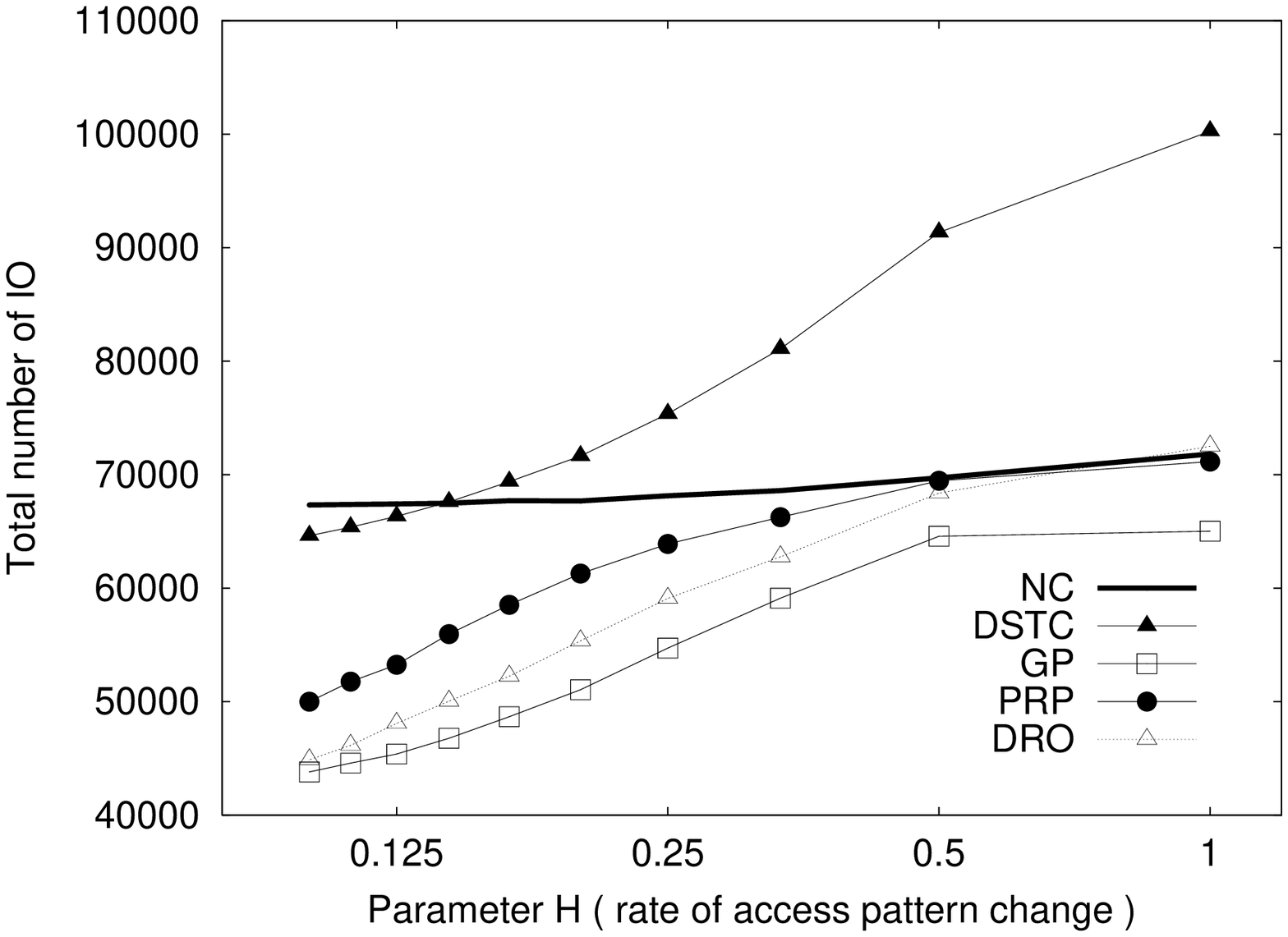}}\quad
  \subfigure[Fenêtre mobile graduelle]{\includegraphics[width = 5.8 cm]{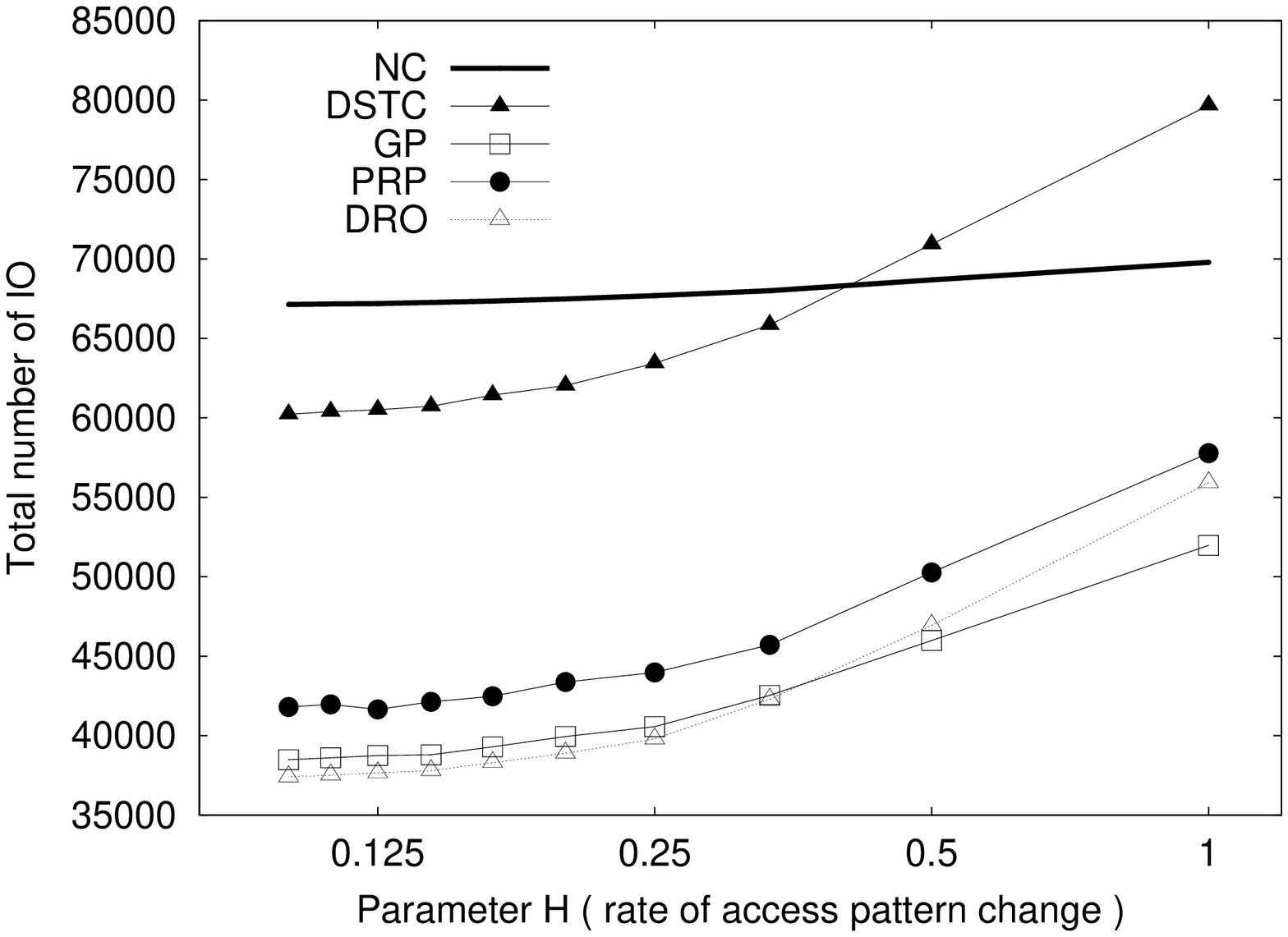}}\quad
  \caption{Résultat de l'intégration de protocoles régionaux et du protocole de dépendance par S-référence}
  \label{res2}
  \end{center}
\end{figure*}

Les résultats que nous avons obtenus sont représentés dans la Figure~\ref{res2}. Lorsque le protocole
\emph{fenêtre mobile} est appliqué (Figure~\ref{res2} (a)), DRO, GP et PRP se montrent à nouveau plus
robustes que DSTC. Cependant, contrairement à ce qui se produisait dans l'expérience précédente,
leurs performances ne deviennent jamais pires que lorsque qu'aucun regroupement
n'est effectué, même quand la séquence d'accès \linebreak change à chaque transaction ($H=1$). Cela s'explique 
par le fait que la sélection par S-référence induit des évolutions de séquence d'accès moins brutales
que le protocole \emph{fenêtre mobile} seul. Dans le cas du protocole
\emph{fenêtre mobile graduelle} (Figure~\ref{res2} (b)), l'écart de performance entre DSTC et les
autres algorithmes demeure le même lorsque la vitesse de changement des séquences d'accès augmente.
Cela indique que DSTC est suffisament robuste pour absorber ce type d'évolution très lente (la seule propriété
qui change est le refroidissement et le réchauffement lent des S-références).

\subsection{Gestionnaires d'objets persistants}

Dans cette série d'expériences, nous avons utilisé DOEF pour comparer les performances de deux 
gestionnaires d'objets persistants existants : SHORE~\cite{shore} et Platypus~\cite{platypus}.

SHORE a été conçu pour répondre de façon efficace aux besoins de nombreux types d'applications,
y compris les langages de programmation orientés-objets. Il s'appuie sur un modèle distribué de
type pair à pair et a été spécifiquement conçu pour être performant. L'architecture en couches de SHORE permet
aux utilisateurs de sélectionner le niveau de service souhaité pour une application particulière.
La couche la plus basse, le SSM (\emph{SHORE Storage Manager}), fournit les primitives de base
pour accéder aux objets persistants.  La politique de gestion de cache de SHORE est CLOCK. Nous avons utilisé la version 2.0 
de SHORE dans tous nos tests.

Platypus est également un gestionnaire d'objets persistants conçu pour offrir les meilleurs temps
d'accès possibles. Il peut fonctionner selon trois modèles distribués : centralisé, client-serveur ou 
client-pair et comprend différents services typiques des SGBDO tels que la journalisation, la
reprise sur panne, un ramasse-miettes et la possibilité d'intégrer des algorithmes de regroupement.
La politique de gestion de cache de Platypus est LRU.

Dans ces expériences, nous avons voulu tester les éléments "de base" entrant en compte dans les
performances de SHORE et Platypus. Aussi n'avons-nous pas intégré de stratégie de regroupement
dans ces deux systèmes.
Nos tests ont été effectués sur une station de travail Solaris~7 dotée de deux processeurs Celeron à 433~MHz,
de 512~Mo de mémoire vive et d'un disque dur de 4~Go. 
A partir du SSM de SHORE, nous avons construit une interface
PSI~\cite{blackburn98c} baptisée PSI-SSM qui nous a permis d'utiliser le même code de DOEF pour SHORE
et Platypus.
Le paramétrage de DOEF et OCB est le même que dans nos expériences sur le regroupement (Tableau~\ref{param_tests}),
à l'exception de la taille de la base qui compte 400000 objets dont la taille varie entre 50 et 1200 octets ---
soit une taille moyenne de 269 octets et une taille totale de la base de 108~Mo. La taille du cache
de SHORE et Platypus a été fixée à 61~Mo.

\begin{figure*}[hbt]
  \begin{center}
  \includegraphics[width = 6 cm]{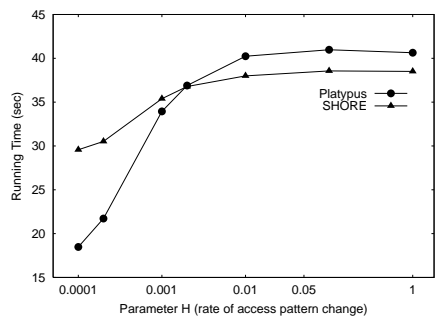}
  \caption{Résultat de l'application du protocole \emph{fenêtre mobile}}
  \label{res3}
  \end{center}
\end{figure*}

Nous avons appliqué le protocole \emph{fenêtre mobile} pour comparer les effets d'évolutions de
séquence d'accès sur les performances de Platypus et SHORE. Les résultats que nous avons
obtenus (Figure~\ref{res3}) montrent que Platypus présente de meilleures performances
que SHORE lorsque la vitesse d'évolution des séquences d'accès (paramètre $H$) est faible,
mais que l'écart se réduit nettement lorsque cette vitesse augmente. Cela s'explique par l'évolution de la localité d'accès.
Lorsque la vitesse de changement des séquences d'accès est basse, la localité d'accès est
élevée (la région chaude est petite et se déplace lentement). Les objets les plus demandés se
trouvent donc en général dans le cache. En revanche, quand les séquences d'accès changent plus
rapidement, la localité d'accès diminue et les accès disques deviennent les plus fréquents.
Nous attribuons ce phénomène à une déficience de Platypus en termes de pagination disque, en raison de
la granularité de verrouillage trop élevée entre son serveur de pages et les processus clients,
qui entraîne un faible degré de concurrence.

\section{Conclusion et perspectives}
\label{conclu}

Nous avons présenté dans cet article les spécifications d'une nouvelle plate-forme d'évaluation des
performances, DOEF, qui permet aux concepteurs et aux utilisateurs de SGBDO de tester un système donné
à l'aide d'une charge dite dynamique. C'est là l'originalité de notre proposition, car si la plupart
des applications réelles présentent des évolutions dans leurs séquences d'accès aux données, 
les bancs d'essais actuels ne modélisent pas ce type de comportement.

Nous avons conçu DOEF comme une plate-forme ouverte et extensible, selon deux axes. Premièrement, comme
c'est à notre connaissance la première tentative d'étudier le comportement dynamique des SGBDO, nous avons
pris grand soin de garantir l'intégration facile de nouveaux modèles d'évolution de séquence d'accès,
principalement grâce à la définition des H-régions. Nous encourageons les chercheurs à utiliser et à
enrichir cette plate-forme pour tester leurs propres idées. Le code utilisé dans nos expériences de
simulation et nos implémentations sur Platypus et SHORE est par ailleurs librement disponible en 
ligne\footnote{\noindent {\tt http://eric.univ-lyon2.fr/$\sim$jdarmont/download/docb-voodb.tar.gz}\\
{\tt http://eric.univ-lyon2.fr/$\sim$jdarmont/download/docb-platypus-shore.tar.gz}}. 

Deuxièmement, bien que nous ayons employé un environnement purement orienté-objets dans cette première
étude, nous pourrions appliquer les concepts développés dans cet article aux bases de données
relationnelles-objets. En effet, puisqu'OCB peut être assez aisément adapté au modèle relationnel-objet
(bien que des extensions soient clairement nécessaires, comme les types de données abstraits et les
tables emboitées, par exemple), DOEF, qui est une surcouche d'OCB, peut également être utilisé dans
un contexte relationnel-objet.

L'objectif principal de DOEF est de permettre aux chercheurs et aux ingénieurs d'observer les 
performances des SGBDO (identifier les composants qui forment des goulots d'étranglement, par
exemple) lorsque les séquences d'accès aux données varient. Les résultats de nos expérimentations
sur les algorithmes de regroupement dynamiques et les gestionnaires d'objets persistants ont démontré
que DOEF permettait d'atteindre cet objectif. Dans le cas des algorithmes de regroupement, nous avons
mis deux choses en évidence. Ces techniques peuvent s'adapter à des évolutions lentes des séquences
d'accès, mais leurs performances s'effondrent lorsque les changements sont rapides. De plus, un
regroupement dit prudent et flexible est indispensable pour prendre en compte de telles évolutions.
En ce qui concerne Platypus et SHORE, l'utilisation de DOEF a permis de mettre en évidence les
problèmes de pagination disque de Platypus.

Ce travail de recherche ouvre plusieurs perspectives. La première concerne évidemment l'exploitation
de DOEF, de manière à continuer d'accumuler de l'expertise et des connaissance sur le comportement
dynamique des SGBDO. De plus, comparer le comportement dynamique de différents systèmes, qui constitue
déjà une tâche intéressante en soi, pourrait nous permettre d'identifier de nouveaux modèles d'évolution
de séquence d'accès à inclure dans DOEF, puisque nous n'avons certainement pas pris en compte tous les
cas possibles. Améliorer ou affiner la définition des H-régions pour prendre en compte la structure du
graphe d'objets pourrait également enrichir DOEF.

Par ailleurs, l'adaptation de DOEF au modèle relationnel-objet se révèle indispensable pour pouvoir 
tester et comparer les performances de ces systèmes et tenter d'identifier leurs goulots d'étranglement. 
Puisque le schéma d'OCB peut être implémenté directement dans un système relationnel-objet, cela reviendrait
à adapter les opérations d'OCB et à en ajouter de nouvelles, spécifiques ou pertinentes dans ce contexte.

Finalement, la capacité de DOEF à évaluer d'autres aspects des performances des SGBDO pourrait
être explorée. Le regroupement est en effet une technique efficace, mais d'autres stratégies comme
la gestion de cache et le préchargement, ainsi que leur utilisation conjointe, pourraient également faire l'objet d'une étude.

\bibliography{doef_bda}

\acknowledgements{Les auteurs remercient Stephen M. Blackburn pour ses commentaires et ses suggestions pertinents lors
de nos expérimentations et de la rédaction de cet article.}

\end{document}